\documentclass[final,1p,times]{elsarticle}

\usepackage{amssymb}
\usepackage{amsmath}
\usepackage{mathtools}

\journal{Nuclear Physics B}

\begin{document}

\begin{frontmatter}



\title{Physics Based \& Machine Learning Methods For Uncertainty Estimation In Turbulence Modeling}


\author[inst1]{Minghan Chu}

\affiliation[inst1]{organization={Department of Mechanical and Materials Engineering, Queen's University},
            city={Kingston},
            state={Ontario},
            country={Canada}}



\begin{abstract}
Turbulent flows play an important role in many scientific and technological design problems. Both Sub-Grid Scale (SGS) models in Large Eddy Simulations (LES) and Reynolds Averaged Navier Stokes (RANS) based modeling will require turbulence models for computational research of turbulent flows in the future. Turbulence model-based simulations suffer from a multitude of causes of forecast uncertainty. For example, the simplifications and assumptions employed to make these turbulence models computationally tractable and economical lead to predictive uncertainty. For safety-critical engineering design applications, we need reliable estimates of this uncertainty. This article focuses on Uncertainty Quantification (UQ) for Computational Fluid Dynamics (CFD) simulations. We review recent advances in the estimate of many types of uncertainty components, including numerical, aleatoric, and epistemic. We go into further depth on the possible use of Machine Learning (ML) methods to quantify these uncertainties. Above all, we elaborate further on significant limitations in these techniques. These range from realizability constraints on the Eigenspace Perturbation Method (EPM) to the requirement for Monte Carlo (MC) approaches for mixed uncertainty. Based on this study, we pinpoint important problems that need to be addressed and offer focused solutions to move beyond these obstacles. 
\end{abstract}




\end{frontmatter}


\section{Introduction }
\label{sec:sample1}

Computational fluid dynamics (CFD) simulations have been widely used to study turbulent flows as they are frequently encountered in both academic and engineering design applications \cite{ferziger2019computational, ferziger2002computational, ferziger1996further}. For Direct Numerical Simulation (DNS) to capture the shortest Kolmogorov length scale of turbulent motion, an extremely fine mesh is required \cite{moin1998direct, moser1999direct}. Consequently, DNS is not feasible for engineering applications due to its significant computational resource demands. Current computational limits restrict DNS to small sections of an airplane wing at high Reynolds numbers. With advances in research, processing overheads are being reduced for large-eddy simulation (LES), which resolves large-scale eddies while modeling small-scale eddies \cite{piomelli1999large}. However, at present, LES is confined to small-scale research investigations for flows with very high Reynolds numbers and intricate geometric features. As a result, Reynolds-averaged Navier-Stokes (RANS) based turbulence models have become popular due to their significantly reduced processing overheads by modeling all scales of turbulent motion \cite{durbin2002perspective, speziale1998turbulence}. Unlike DNS and LES, which aim to capture the highly accurate physics of turbulent flows, the RANS technique describes both higher-order and lower-order properties using simplifying modeling assumptions. Consequently, RANS remains the most commonly used CFD method in engineering applications. However, these simplified assumptions introduce sources of uncertainty in CFD solutions, such as assumptions made in model formulation, errors in the initial and boundary conditions, and inadequate mesh resolutions.

Aleatory and epistemic sources of error or uncertainty are generally classified as the key contributors to predictive uncertainty \cite{smith2013uncertainty, stern1999verification, peter2010verification, giles1998adjoint, barth2016overview}. Uncertainty quantification (UQ) seeks to evaluate the influence of propagated uncertainties on output predictions and then minimize these uncertainties. This process requires a suitable definition of uncertainty, often expressed as a probability distribution. Aleatory uncertainties arise from a system's inherent variability (imprecision) and inputs, such as initial and boundary conditions \cite{duraisamy2019turbulence}. These uncertainties can be caused by discrepancies between simulations and reality in terms of initial or boundary conditions, inaccuracies in measuring geometry and material characteristics, and other factors. Aleatory uncertainties are intrinsically stochastic (usually defined probabilistically), irreducible, and unbiased. They can lead to exponentially increasing complexity in simulations of real systems. However, by reconstructing stochastic terms, aleatory uncertainties can be minimized, leading to more precise initial and boundary conditions (priors). Throughout the simulation, the intrinsic variance of aleatory uncertainties may propagate. Several UQ techniques have been developed to measure and reduce aleatory uncertainties. Bayesian inference methods such as Polynomial Chaos \cite{najm2009uncertainty} and Stochastic Collocation \cite{mathelin2003stochastic} assign posterior probability distributions to model parameters through a calibration process. For example, model coefficients may be represented as random variables \cite{loeven2008airfoil, ahlfeld2017single}, flow domains may be represented as random fields \cite{dow2015implications, doostan2016bi}, and boundary and initial conditions may also be represented as random \cite{pecnik2011assessment}.

Conversely, model inadequacy—the intrinsic incapacity of models to adequately capture the physics of turbulence—causes epistemic uncertainties. This type of reducible/biased uncertainty will persist even if aleatory uncertainties are reduced. Epistemic uncertainties include structural uncertainty arising from the constraints of the model expression and uncertainty related to a turbulence model's coefficients. Under the Bayesian framework, both aleatory and epistemic uncertainties can be represented by probability distributions, provided there is sufficient prior knowledge to construct one \cite{najm2009uncertainty}. In daily engineering modeling of turbulent flows, structural uncertainty can be a dominant source of uncertainty \cite{duraisamy2017status}. Employing high-fidelity models can help reduce epistemic uncertainty, although these models often come with prohibitive processing costs. To benefit from characterizing structural uncertainties while maintaining reduced computational costs, it is worthwhile to use an eddy-viscosity model based on Reynolds-Averaged Navier-Stokes (RANS). The Eigenspace Perturbation Method (EPM) \cite{emory2013modeling,iaccarino2017eigenspace} is currently the only physics-based method available for estimating uncertainty resulting from turbulence models. In the EPM, we perturb the spectrum decomposition components of the expected Reynolds stress tensor to quantify the sensitivity of predictions to structural uncertainties. EPM can estimate the propagating impact of uncertainty on predictions at a relatively low cost, as it does not require a very fine mesh or substantial \textit{a priori} data. This method has been applied to various engineering problems, including the design of urban canopies \cite{gorle2019epistemic}, aerospace design and analysis \cite{mishra2019uncertainty, mishra2017rans, mishra2019estimating, mishra2017uncertainty}, design under uncertainty (DUU) \cite{demir2023robust, cook2019optimization, mishra2020design, righi2023uncertainties, alonso2017scalable}, and virtual certification of aircraft \cite{mukhopadhaya2020multi, nigam2021toolset, mishra2019linear}.


Applications of machine learning (ML)-based models to fluid dynamics and turbulence are growing in popularity \cite{duraisamy2019turbulence, ihme2022combustion, brunton2020machine, chung2021data}. Many researchers have utilized data to create functions that can forecast the discrepancies in turbulence model predictions \cite{xiao2016quantifying, wu2018physics, heyse2021estimating, heyse2021data, zeng2022adaptive}. These studies have focused on estimating epistemic uncertainty or turbulence model form. However, an overfitting tendency is often observed in many of these ML models. Consequently, the ML models become less generalizable, making the trained models accurate only for the flows they were trained on. Due to the sophistication of ML models, a substantial amount of relevant data is required for training. This is not always feasible in engineering design, especially when considering novel designs. Additionally, sophisticated ML models frequently function as black box models, with poorly understood internal workings. This lack of transparency hinders our ability to verify and validate these models, raising concerns about their trustworthiness and applicability in engineering contexts \cite{chung2022interpretable}. While the data is specific to the training examples from which it was gathered, the laws of physics are universally applicable to all flows. Therefore, it is imperative that these ML models for uncertainty quantification incorporate physics-based information.


Beyond aleatory and epistemic uncertainties, numerical and algorithmic uncertainty emerges as a result of discretization or the methods used to solve partial differential equations (PDEs). The Grid Convergence Index (GCI) introduced by Roache \cite{roache1993method, roache1997quantification, roache1998verification} and Richardson Extrapolation (or $h^2$ extrapolation) \cite{richardson1927viii} are commonly employed techniques to assess discretization uncertainty. These methods involve refining the numerical mesh (or grid) to quantify the numerical uncertainty in CFD simulations.


This paper examines various Uncertainty Quantification (UQ) approaches for assessing aleatory and epistemic uncertainties, with a specific emphasis on the model-form uncertainties inherent in eddy-viscosity Reynolds-Averaged Navier-Stokes (RANS) models. The paper is structured as follows: Section 1 provides an overview of the sources of uncertainties in Computational Fluid Dynamics (CFD) simulations and the techniques employed to quantify these uncertainties. Section 2 delves into model-form uncertainties in turbulence models, particularly focusing on Reynolds-Averaged Navier-Stokes models. Section 3 discusses Uncertainty Quantification utilizing forward models. Section 4 addresses Uncertainty Quantification utilizing backward models. Sections 5, 6, and 7 concentrate on the Eigenspace Perturbation Method (EPM) due to its relevance in the field of turbulence modeling. Section 5 covers Eigenvalue perturbation. Section 6 discusses Eigenvector perturbation. Section 7 explores perturbations to turbulent kinetic energy. Section 8 provides a summary of the manuscript and suggests potential avenues for future research.

\section{Model form uncertainty in CFD simulations with turbulence models}
In relation to categories, predictive uncertainty can be classified as either aleatory or epistemic. Epistemic uncertainty arises from a lack of knowledge in describing a quantity, which can be mitigated by acquiring more knowledge about the physics of flow. Epistemic uncertainty can be further divided into parametric and non-parametric uncertainties (structural) based on where uncertainties are introduced. Parametric uncertainties are introduced to the coefficients of the closure model, resulting in less understanding of the underlying physics. On the other hand, non-parametric uncertainties are inherent to the modeled terms and provide valuable insights into the physical aspects, such as the eddy viscosity \cite{wang2010quantification}, the source terms of transport equations, or the Reynolds stress \cite{xiao2016quantifying}.

From a probabilistic standpoint, uncertain quantities of interest can be categorized as either parametric or non-parametric random variables. The classification of these random variables depends on how they are indexed. For example, a scalar random variable ($X$) can be represented as a vector of random variables $\boldsymbol{X}=\left[X_1, \cdots, X_n\right]$, where each element is indexed by integers. On the other hand, a random field $X(y)$ refers to a collection of random variables indexed by the spatial coordinate $y$. Additionally, it can also be interpreted as a collection of stochastic variables indexed by the time coordinate $t$.

Based on the nature of uncertainty, the forward (data-free) \& the backward (data-driven) approaches are utilized for the quantification and mitigation of uncertainty. Forward methods require predefined probability distributions for both parametric and non-parametric uncertainties, which are then propagated through the governing RANS equations. Conversely, backward methods incorporate observed data to estimate parametric/non-parametric uncertainties. It is important to highlight that the estimated probability distributions are subsequently employed in the prediction phase, similar to the forward approach.

Aleatory uncertainty, on the other hand, arises from incipient and inherent variability. Homoscedastic aleatory uncertainty exhibits consistent variance for a variable, while heteroscedastic uncertainty displays varying variances. From a frequentist probability viewpoint, aleatory uncertainties are inherent in stochastic quantities characterized by probability density functions (PDF). However, PDF cannot be established for quantities with epistemic uncertainty within the frequentist probability framework. This limitation is not present in the Bayesian perspective. In the Bayesian framework, a PDF is constructed based on the level of belief, provided that adequate prior information is accessible.

\section{Uncertainty quantification via forward appoaches}
A known prior probability distribution $p(\theta)$ is required for forward methods, that are used to address uncertainties in model parameters. These methods involve sampling from high probability regions, as specified by the prior distribution \cite{ghanem2003stochastic}. Forward methods can be categorized into spectral methods and Monte Carlo methods \cite{ghanem2003stochastic, glasserman2004monte}. Spectral methods discretize the uncertain space of random variables using orthogonal basis functions. Polynomial Chaos Expansion (PCE) is an example of a spectral method that constructs an approximate relationship between input parameters and model output in the presence of uncertainty. On the other hand, Monte Carlo methods estimate the output uncertainty by considering input random variables. However, Monte Carlo methods have a slow convergence rate, which is proportional to the inverse square root of the number of samples, denoted as $\mathcal{O}\left(N^{-1 / 2}\right)$ \cite{glasserman2004monte}.

\section{Uncertainty Quantification using backward inference}
Frameworks that are Bayesian start with an initial belief about the uncertainty in parameters and then analyze data to update these beliefs regarding parameter uncertainty. The degree of change to the initial beliefs about parameter uncertainty on seeing data is governed by the Bayes Theorem \cite{berkson1930bayes}:
\begin{equation}\label{Eq:Bayes}
    p(\theta \mid Z)=\frac{p(Z \mid \theta) p(\theta)}{p(Z)},
\end{equation}

where $p(\theta)$ is the prior, $p(Z \mid \theta)$ is the likelihood, $p(\theta \mid Z)$ is the posterior probability, and the $p(z)$ is total probability of the observed data for normalization. Equation \ref{Eq:Bayes} states that the posterior probability is proportional to the $p(\theta)$ and the $p(Z \mid \theta)$. The prior probability distribution represents our beliefs about the parameter uncertainty before seeing any data. The likelihood represents the probability of seeing the training data conditioned upon the value of the parameter. This results in the posterior probability which is the final belief about parameter uncertainty after seeing the data. Thus the final uncertainty is based on the initial (or prior) beliefs and on the nature and magnitude of the data.

\subsection{Markov Chain Monte Carlo Based Frameworks}
In opposition to forward methods, which are simple and use standard Monte Carlo sampling, posterior methods rely on Bayes' theorem to sample from an undetermined posterior probability distribution. Because standard Monte Carlo sampling has difficulty finding regions of high posterior probability, standard Monte Carlo is not often used in Bayesian frameworks. As an alternative, Markov chain Monte Carlo (MCMC) methods are often used. MCMC methods belong to the category of sequential sampling techniques, where the next sampling state is based only on the current state. This approach allows sampling to be concentrated in high-probability regions while occasionally exploring low-probability regions (the tail). Using a target distribution, the Markov Chain Monte Carlo algorithm generates samples from that distribution, creating a Markov chain with a stationary distribution corresponding to the desired target distribution.


\begin{figure} 
\centerline{\includegraphics[width=5.5in]{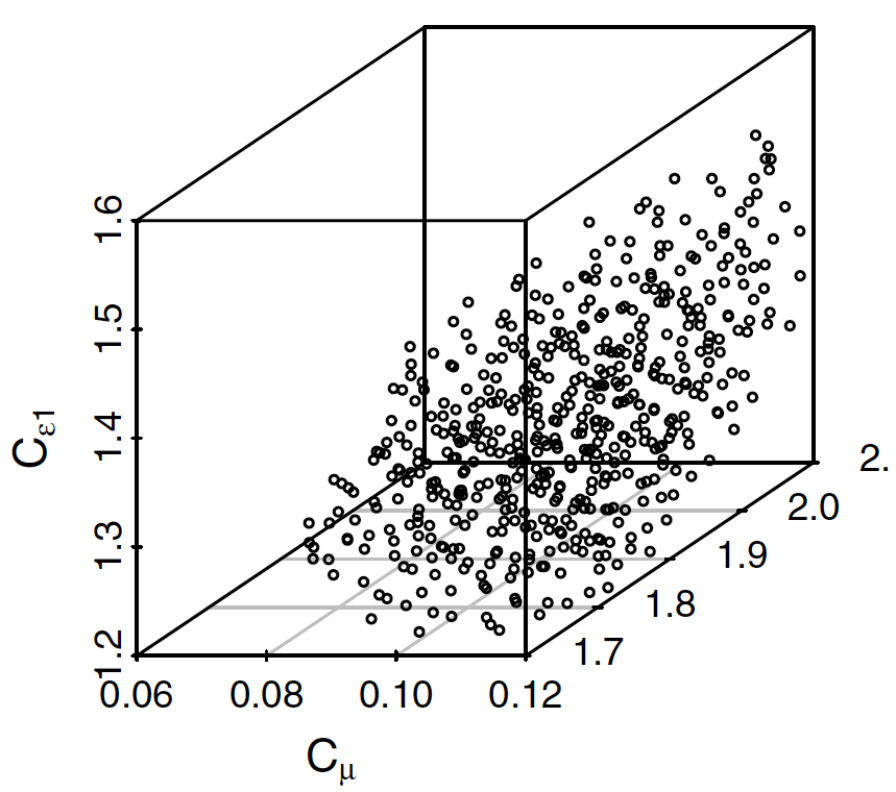}}
\caption{Visualization of the samples in $(C_{\mu}, C_{\epsilon 2}, C_{\epsilon 1})$ space that constitute $\mathcal{R}$ for the RMSE that is below the $20$th percentile.}
\label{fig:MCMC subspace}
\end{figure}

While MCMC is the gold standard of Bayesian inference and posterior sampling, its practical application requires an extremely large number of samples to reach statistical convergence. Usually, the number of required samples ranges from $\mathcal{O}(10^{5})$ to $\mathcal{O}(10^{6})$ , which varies depending on the shape of the posterior, the dimensionality of the problem, the prior used, the efficiency of the sampling, etc. In Computational Fluid Dynamics (CFD), each evaluation or sample involves simulations that may take hours, days or even weeks depending on the complexity of the flow and the fidelity of the required samples.

We may therefore infer that it is not possible to do RANS simulations for each likelihood evaluation within the MCMC sampling. This is not just because of the large number of samples required, but also because of the sequential nature of standard MCMC methods. The suggestion of the next sample in these algorithms depends on the evaluation of the posterior of the present state, which makes it more difficult to carry out complete simulations at every stage. Surrogate models are often used in MCMC-based model uncertainty quantification as a solution to this problem. They are used to reduce the significant computing costs related to RANS simulations for assessing probability.

An RANS model is calibrated using a Bayesian calibration technique \cite{ray2016bayesian,ray2018learning}, and its model parameters, $\mathbf{C}=(C_{\mu}, C_{\epsilon 2}, C_{\epsilon 1})$, are estimated. The MCMC method is used to solve the Bayesian approach. $P\left(\boldsymbol{C}, \sigma^2 \mid \boldsymbol{y}_e\right)$ is the joint probability density function of the parameters and the model-data misfit, and it is conditional on the observed data $y_{e}$. Specifically, $\Pi_{1}(\boldsymbol{C})$ and $\Pi_{2}(\sigma^{2})$ indicate the previous views about the distribution of $\boldsymbol{C}$ and $\sigma^{2}$. Given a parameter setting $\boldsymbol{C}$, the probability of witnessing $y_{e}$ is written as $\mathcal{L}(y_e \mid \boldsymbol{C})$. This is expanded to:

\begin{equation}\label{Eq:likelihood}
    \mathcal{L}\left(\boldsymbol{y}_e \mid \boldsymbol{C}, \sigma^2\right) \propto \frac{1}{\sigma^{N_p}} \exp \left(-\frac{\left\|\boldsymbol{y}_e-\boldsymbol{y}_m(\boldsymbol{C})\right\|_2^2}{2 \sigma^2}\right)
\end{equation}

Based on the Bayes' theorem in Eqn. \ref{Eq:Bayes}, the posterior distribution of $(\boldsymbol{C}, \sigma^{2})$ is defined as:

\begin{equation}\label{Eq:posterior}
    \begin{aligned} P\left(\boldsymbol{C}, \sigma^2 \mid \boldsymbol{y}_e\right) & \propto \mathcal{L}\left(\boldsymbol{y}_e \mid \boldsymbol{C}, \sigma^2\right) \Pi_1(\boldsymbol{C}) \Pi_2\left(\sigma^2\right) \\ & \propto \frac{1}{\sigma^{N_p}} \exp \left(-\frac{\left\|\boldsymbol{y}_e-\boldsymbol{y}_m(\boldsymbol{C})\right\|_2^2}{2 \sigma^2}\right) \Pi_1(\boldsymbol{C}) \Pi_2\left(\sigma^2\right).\end{aligned}
\end{equation}

To rebuild the posterior probability distribution $P\left(\boldsymbol{C}, \sigma^2 \mid \boldsymbol{y}_e\right)$, samples ($O(10^{4})$) of $\{\boldsymbol{C}, \sigma^{2} \} $ can be drawn using the MCMC technique. Using histograms or Kernel density estimates, $P\left(\boldsymbol{C}, \sigma^2 \mid \boldsymbol{y}_e\right)$ \cite{silverman2018density}. A sufficient number of samples of $\{\boldsymbol{C}, \sigma^{2} \} $ are needed when the MCMC chain converges to a stable posterior probability distribution, and an algorithm is needed to assess the sufficiency, e.g., \cite{raftery1996implementing}. 

The MCMC technique is unfeasible due to the enormous number of samples and therefore high computing cost. Specifically, each of the $O(10^{4})$ samples requires a RANS simulation in order to give $\mathbf{y_{m}(C)}$ in Eqn. \ref{Eq:posterior}. Therefore, by mapping the dependency of the unknown values of interest on $\mathbf{C}$, a polynomial surrogate may be used in place of a RANS model to drastically minimize the computing cost. We build the (polynomial) surrogate model using $\boldsymbol{C}$, that is, $\boldsymbol{C \in R}$. Keep in mind that the surrogate approach can only be applied to state spaces with smaller dimensions. This mapping holds true in the support of $\Pi_1(\boldsymbol{C})$ within a reasonable error bound.  Generally, the boundaries of the uncertain parameter space $\mathbf{C}$ may be determined; nevertheless, random selections of parameter combinations from this space may be physically impractical. RANS simulations could crash as a result of it. Consequently, we must choose a region of $\mathbf{R}$ inside $\mathbf{C}$ space that has the values that result in flowfields that are physically plausible. Therefore, in order to limit the values of parameters to a Region $\mathcal{R}$, the creation of an informative prior is necessary. The subset $\boldsymbol{R}$ space is found to exclude a significant amount of $\boldsymbol{C}$, as seen in Figure \ref{fig:MCMC subspace} \cite{ray2016bayesian}. The $\boldsymbol{R}$ subspace is determined by comparing the root-mean-square error (RMSE) between the vorticities from each simulation and the experimental data on the crossplane \cite{ray2016bayesian}. The values of $(C_{\mu}, C_{\epsilon 2}, C_{\epsilon 1})$ that resulted in an RMSE below the $20th$ percentile among the chosen runs define $\boldsymbol{R}$, as illustrated in Figure \ref{fig:MCMC subspace}.






\subsection{Using the Ensemble Kalman Filter for Approximate Bayesian Inference}
The posterior distribution can be sampled with extreme precision using the MCMC method, although a significant number of samples are needed. Alternative approximation Bayesian inference procedures are useful when only lower-order statistical moments, such as the mean and variance, are significant and exact probability is not. Instead of capturing the full posterior distribution, these methods identify the mode (peak) of the posterior distribution using the maximum a posteriori (MAP) probability estimate. In this paper, we focus on the ensemble Kalman filter approach, a particular MAP technique that has been extensively applied in Bayesian contexts \cite{iglesias2013ensemble,xiao2016quantifying}. The model is integrated forward in time using the sequential filter approach of the original Ensemble Kalman Filter (EnKF). The model is reinitialized before further integration takes place, whenever the observed data become available, by comparing the predictions with the observed data. The initial EnKF technique has been extensively utilized in numerous applications of data assimilation for state and parameter estimation, particularly in the fields of weather forecasting \cite{houtekamer2001sequential}, reservoir modeling \cite{aanonsen2009ensemble}, and oceanography \cite{evensen1996assimilation}.

A more modern iterative EnKF (IEnKF) is based on the original EnKF and entails repeatedly performing the EnKF update process. Following the first EnKF update, IEnKF runs further iterations, refining the estimation with each iteration by continuously updating the ensemble and assimilating observations. This iterative procedure aids in enhancing the state estimation's accuracy, particularly in circumstances when the original EnKF may show flaws or limitations. The system state is first extended to include the physical states $x(t)$, such as the kinetic energy fields of turbulence, pressure, and/or velocity, as well as the unobservable parameters $\mathbf{\theta}$, such as the viscosity field or model coefficients, which need to be inferred. We call this kind of system state an augmented state. In essence, ensemble Kalman-based techniques use the Kalman formulation to update an ensemble of an augmented state by fusing observed data at a certain moment with the model's forecast \cite{evensen2003ensemble,evensen2009data}. 

To infer unknown $u$ using IEnKF, we need observed data as

\begin{equation} \label{Eq:Observations}
    y = \mathcal{G}(u) + \epsilon,
\end{equation}

where the forward response operator that maps $u$ to the observations space is denoted by $\mathcal{G}$. The forward response $\mathcal{G}$ from $u$ is calculated by a PDE system that describes a physical system, as you can see. It is assumed that $\epsilon$ has a normal distribution with a mean of zero and a known covariance, $Cov$. Let $X$ and $Y$ be the prediction space and the observation space, respectively, to solve the inverse issue. An augmented state's artificial dynamics can be constructed as

\begin{equation} \label{Eq:AugmentStateMapping}
    \Xi\left(z\right)=\begin{pmatrix} y\\ \mathcal{G}(u). \end{pmatrix} \quad \text{for} \quad z = \begin{pmatrix}
        u\\ p 
    \end{pmatrix} \in Z.
\end{equation}

In Eqn. \ref{Eq:AugmentStateMapping}, $z$ is the augmented state being mapped to $Z$ space. Therefore, the artificial dynamics can be expressed as 

\begin{equation}\label{Eq:airtificaldynamics_AugState}
    z_{n+1} = \Xi\left( z_{n} \right).
\end{equation}

The observed data in Eqn. \ref{Eq:Observations} can be assumed based on the artificial dynamics to be

\begin{equation} \label{Eq:y_n+1}
    y_{n+1} = Hz_{n+1} + \epsilon_{n+1},
\end{equation}

where the projection operator or measurement function mapping $Z$ space to $Y$ space is denoted by $H = (0, I)$. Subsequently, the goal of the EnKF method is to calculate the unknown $u$ in Eqn. \ref{Eq:Observations} by estimating the augmented state in Eqn. \ref{Eq:airtificaldynamics_AugState}. The augmented state (Eqn. \ref{Eq:airtificaldynamics_AugState}) and the observed data (Eqn. \ref{Eq:Observations}) are combined to update an ensemble of particles at each iteration. In Eqns. \ref{Eq:airtificaldynamics_AugState} and \ref{Eq:y_n+1}, the iteration index $n$ denotes an artificial time, whereas the real time is associated with $\mathcal{G}$ and is independent of $n$. 

Since the inverse problems from Eqns. \ref{Eq:Observations} to \ref{Eq:y_n+1} are ill-posed, regularization is necessary by taking into account past knowledge of $u$ in a finite-dimensional subspace $\mathcal{A}$, that is, $\mathcal{A} \in X$. $\mathcal{A}$ still contains the EnKF-based solution from Eqn. \ref{Eq:Observations}. In order to estimate the unknown $u$, the EnFK framework creates an interacting ensemble of particles $\{z^{(j)}_{n}\}^{J}_{j=1}$ using:

\begin{equation}\label{Eq:unknownu}
    u_{n} \equiv \frac{1}{J}  \sum_{j=1}^{J} u^{(j+1)}_{n} = \frac{1}{J}  \sum_{j=1}^{J} H^{\perp}z^{(j)}_{n}.
\end{equation}

The EnKF method requires a first guess of $\{z^{(j)}_{0}\}^{J}_{j=1}$ to be initiated. The ensemble $\{z^{(j)}_{0}\}^{J}_{j=1}$ can be created by forming an ensemble $\{\psi^{(j)}\}^{J}_{j=1}$ within the $\mathcal{A}$ space where the solution of the unknown $u$ is sought. Then, we can set

\begin{equation} \label{Eq:z_{0}}
    z^{(j)}_{0} = \begin{pmatrix} \psi^{(j)} \\ \mathcal{G}(\psi^{(j)}). \end{pmatrix}.
\end{equation}

The mean of the first ensemble in $\mathcal{A}$ space is the corresponding $u_{0}$ (see to Eqn. \ref{Eq:unknownu}). With the available prior knowledge, such as Gaussian $N(\overline{u}, P)$ and $\overline{u} + \sqrt{\lambda_{j}}\phi_{j}$, the initial ensemble $\{\psi^{(j)}\}^{J}_{j=1}$ can be constructed. Here, $(\lambda_{j},\phi_{j})$ represents the eigenvalue and eigenvector Paris of $P$ with eigenvalue in descending order, i.e., the Karhunen-Alfred Lo\'eve (KL) basis.

The ensemble of particles $\{z^{(j)}_{n}\}^{J}_{j=1}$ will be updated iteratively by the EnKF technique with $\{z^{(j)}_{0}\}^{J}_{j=1}$ specified. The prediction step and the update step at each iteration make up the overall structure of the iterative EnKF. The ensemble of particles is propagated in the prediction stage using Eqn. \ref{Eq:airtificaldynamics_AugState}. The ensemble that is being mapped to the observed data space brings the knowledge from the forward model because the prediction phase uses a forward model. The prediction step is defined as:

\begin{equation} \label{Eq:ensemble_AugState}
    \widehat{z}^{(j)}_{n+1} = \Xi\left( z^{j}_{n} \right).
\end{equation}

From Eqn. \ref{Eq:ensemble_AugState}, the augmented state $\widehat{z}^{(j)}_{n+1}$ can be defined as its mean and covariance as follows:

\begin{equation}
    \hat{z}_{n+1} = \frac{1}{J}  \sum_{j=1}^{J} \widehat{z}^{(j)}_{n+1} 
\end{equation}

\begin{equation}
    P_{n+1} = \frac{1}{J} \sum_{j=1}^{J} \widehat{z}^{(j)}_{n+1}(\widehat{z}^{(j)}_{n+1})^{T} - \overline{z}_{n+1}\overline{z}^{T}_{n+1}.
\end{equation}

In the update step, the calculation of the Kalman gain blending covariance matrices from the prediction and the observed data is defined as follows: 

\begin{equation}
    K_{n+1} = P_{n+1}H^{\star}(HP_{n+1}H^{\star}+R)^{-1},
\end{equation}

where $H^{\star}$ is the adjoint operator of $H$. With Kalman gain, the updated augmented state for each ensemble member can be computed as follows:

\begin{equation}
    z_{n+1}^{(j)} = I\widehat{z}^{(j)}_{n+1} + K_{n+1}(y^{(j)}_{n+1} - H\widehat{z}^{(j)}_{n+1})
\end{equation}

where $y^{(j)}_{n+1} - H\widehat{z}^{(j)}_{n+1}$ is the residual, and 

\begin{equation}
    y^{(j)}_{n+1} = y + \epsilon^{(j)}_{n+1}. 
\end{equation}

The update stage compares the mapped ensemble with the noisy observed data in the observed data space, trying to adjust the ensemble to better fit the observed data. From Eqn. \ref{Eq:unknownu}, the mean of the unknown $u$ is updated as follows:

\begin{equation}\label{Eq:unknownu_update}
    u_{n+1} \equiv \frac{1}{J}  \sum_{j=1}^{J} u^{(j+1)}_{n+1} = \frac{1}{J}  \sum_{j=1}^{J} H^{\perp}z^{(j)}_{n+1}.
\end{equation}

The stopping criterion for the EnKF method is based on the discrepancy principle, from which the EnKF method is terminated for the first $n$ such that 

\begin{equation}\label{Eq:convergence}
    \left\Vert  y- \mathcal{G}(u_{n}) \right\Vert_{R} \leq \tau \left\Vert \epsilon^{\star} \right\Vert_{R} \quad\quad \text{for some $\tau > 1$},
\end{equation}

where $\epsilon^{\star}$ is the noise in the true observed data.

\subsection{Physics-informed IEnKF in Bayesian framework}
\begin{figure} 
\centerline{\includegraphics[width=5.5in]{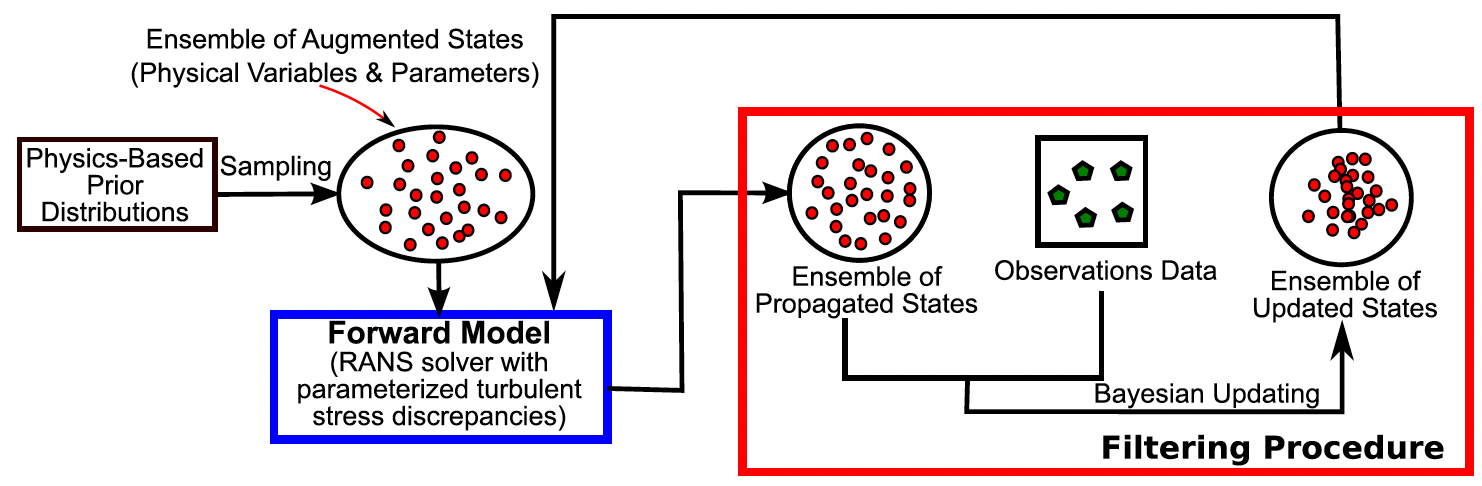}}
\caption{Schematic outlined of the IEnKF algorithm for approximate Bayesian inference.}
\label{fig:XiaoBayesianFramework.png}
\end{figure}

The IEnKF approach uses a Bayesian framework to integrate observable data with prior information based on physics. The goal of this procedure is to update an enhanced state's posterior distributions. Fig. \ref{fig:XiaoBayesianFramework.png} illustrates this iterative plan. The augmented state in this instance is $\mathbf{X} = [\mathbf{u}, \tau^{param}]^{T}$, where the Reynolds stress field is denoted by $\tau$, the velocity field is represented by $\mathbf{u}$, and parameterized is indicated by $param$. By contrasting the baseline prediction for velocity fields with the available observed data, the objective is to estimate or rebuild Reynolds stresses. The initial prior ensemble of augmented states ${\mathbf{u}}^{N}_{j=1}$ and ${\mathbf{\tau^{param}}}^{N}_{j=1}$ are generated from sampling the baseline RANS prediction \cite{xiao2016quantifying}, where $\mathbf{N}$ is the sample size. The physics-based priors give $\tau^{param}$ a physical meaning. Keep in mind that this baseline simulation is only run once for basic setup. An ensemble of ${\mathbf{\tau^{param}}}^{N}_{j=1}$ is included in the forward model. The corresponding ${\mathbf{u}}^{N}_{j=1}$ for each sample in the ensemble ${\mathbf{\tau^{param}}}^{N}_{j=1}$ is obtained by solving the RANS equations. The Kalman filtering technique (also known as the Kalman update step) compares the available observed velocity data with the ensemble mean of ${\mathbf{u}}^{N}_{j=1}$. The ensemble augmented states ${\mathbf{x}}^{N}_{j=1}$ are changed as a consequence. A stopping criterion is based on the Eqn. \ref{Eq:convergence} when statistical convergence of the ensemble is achieved.

RANS equations (PDEs) are considered to be correct in the case presented in Fig. \ref{fig:XiaoBayesianFramework.png}; however, they contain unknown Reynolds stress fields, which cannot be solved directly and instead need to be estimated using a turbulence model. In RANS modeling, the Reynolds stress field is thought to be the main source of uncertainty \cite{oliver2009uncertainty}. Consequently, it is possible to refer to these uncertainty Reynolds stress fields as latent physical fields \cite{strofer2020enforcing}. Latent physical fields encompass not only the initial circumstances of the fields of interest but also any physical fields associated with the fields of interest and constant physical qualities like density or viscosity fields. It is possible to deduce latent physical fields from observable data in the relevant field. 

\begin{equation}
    \mathcal{M}(u; l) = 0
\end{equation}

where $\mathcal{M}$ represents the governing PDEs of a dynamic system, $u$ represents fields of interest, and $l$ represents latent fields. 

The fields of interest can be related to the latent fields as

\begin{equation}
    u = \mathcal{F}(l).
\end{equation}

The only source of uncertainty is latent fields because the PDEs are thought to be accurate. As such, the latent uncertain fields can be inferred to improve the anticipated fields of interest. Utilizing sparse observations, numerous research have been carried out to infer the latent uncertain fields \cite{kennedy2001bayesian, oliver2015validating,duraisamy2019turbulence,xiao2019quantification}. A key factor influencing Bayesian inference is the choice of statistical model defining the prior distribution, particularly its covariance kernel. In the example shown in Fig. \ref{fig:XiaoBayesianFramework.png} \cite{xiao2016quantifying}, the prior for the latent Reynolds stress field is modeled as Gaussian process: $\mathcal{G}\mathcal{P}(0, k)$, where

\begin{equation}
    K(x,x^{'}) = \sigma(x)\sigma(x^{'})exp(-\frac{|x-x^{'}|^{2}}{L^{2}})
\end{equation}

is the covariance kernel for two locations $x$ and $x^{'}$. Large disparities in specific domain regions are reflected in the spatial variation of the variance $\sigma(x)$. The correction length shown by $L$ is determined by the local turbulence length scale. Although updated fields of interest may be improved, Bayesian inference issues are ill-posed, and the Gaussian process may not be able to infer the true latent field. Consequently, a more sophisticated statistical model than a straightforward Gaussian process can be used to impose more physically-realistic global limitations. The representation of the random latent field is chosen so as to guarantee automatic conformity to the physical limitations in any field realization. To enforce a positive constraint, for example, the latent field could be modeled as a lognormal process.

   

\begin{subequations}
\begin{equation}
   \tau = e^{\delta}
\end{equation}    
\begin{equation}
   \delta \sim \mathcal{G}P(log{(\Tilde{\tau}),K}),
\end{equation}
\end{subequations}

where $\Tilde{\tau}$ is the baseline solution. More constraints, particularly regarding covariance kernel can be found in \cite{strofer2020enforcing}.




\subsection{Forward propagation of uncertainty}
Rather than discussing the inference uncertainties involved in the backward propagation, this section concentrates on the uncertainties related to the prediction tasks. Aleatory and epistemic uncertainty are included in the prediction uncertainties. The structure of the utilized turbulence model and its limitations in representing turbulence physics are the source of the epistemic uncertainties. Measurement mistakes in the starting or boundary conditions are the cause of the aleatory uncertainty. 

\subsubsection{Model Form (Structural) uncertainty}
The primary source of uncertainty in CFD simulations of turbulent flows is the model form uncertainty in turbulence models. According to some \cite{zang2002needs}, this is the biggest problem in the design of aerospace vehicles. The Boussinesq turbulent-viscosity hypothesis, which states that anisotropic Reynolds stresses are linearly correlated with the mean rate of strain, is used to approximate Reynolds stress terms in RANS models. The limits of these models—also referred to as linear eddy viscosity models—in managing complex flow scenarios, like those with prominent streamline curvature, separation, and reattachment, are well-documented. Although direct numerical simulations (DNS) and large eddy simulations (LES) provide extremely precise solutions for such complexities, their computing demands—particularly for high Reynolds number flows—often make them unaffordable in terms of both time and computational resources. As a result, evaluating the RANS model's uncertainties becomes a useful substitute for improving prediction abilities in engineering applications. It is only appropriate to take into account more costly LES or DNS simulations if the RANS model's intrinsic uncertainties are greater than acceptable bounds.

RANS modeling is still widely used in many engineering domains because of its reasonable robustness and relatively low processing cost. RANS models, in contrast, require the solution of two equations with two unknowns. In the next step of Reynolds Stress Modeling, eight equations with eight unknowns must be solved. At best, Reynolds Stress Model accuracy is likewise constrained \cite{pope2001turbulent, mishra2016sensitivity}. Methods such as DNS and LES require computational costs that are orders of magnitude larger. RANS models are widely used in this context because they strike a balance between great computational efficiency and high robustness. In order to reduce the computational cost, approximations and simplifications were used in the RANS model development process. This resulted in significant model-form uncertainty in the RANS model predictions.  Model form uncertainty from turbulence models has been quantified by many researchers using various methods. For instance, researchers have employed an ensemble of turbulence models rather than a single model and have estimated the uncertainty of the turbulence model by analyzing the variability in their predictions \cite{vuruskan2019impact, vuruskan2022impact}. More sophisticated turbulence models, such as Reynolds Stress Models (RSM) \cite{pope2001turbulent}, have been used by other researchers. Since all turbulence models have model form uncertainties, this does not eliminate turbulence model uncertainty in addition to increasing computing costs. In order to estimate the model-form uncertainty in RANS modeling, Iaccarino and colleagues \cite{emory2011modeling,iaccarino2017eigenspace,gorle2013framework,emory2013modeling,gorle2014deviation, Emory2014Thesis} broke down the Reynolds stress field and perturbed its magnitude, eigenvalues, and eigenvectors toward their limiting states within the range that is physically realizable. 

\begin{figure} 
\centerline{\includegraphics[width=5.5in]{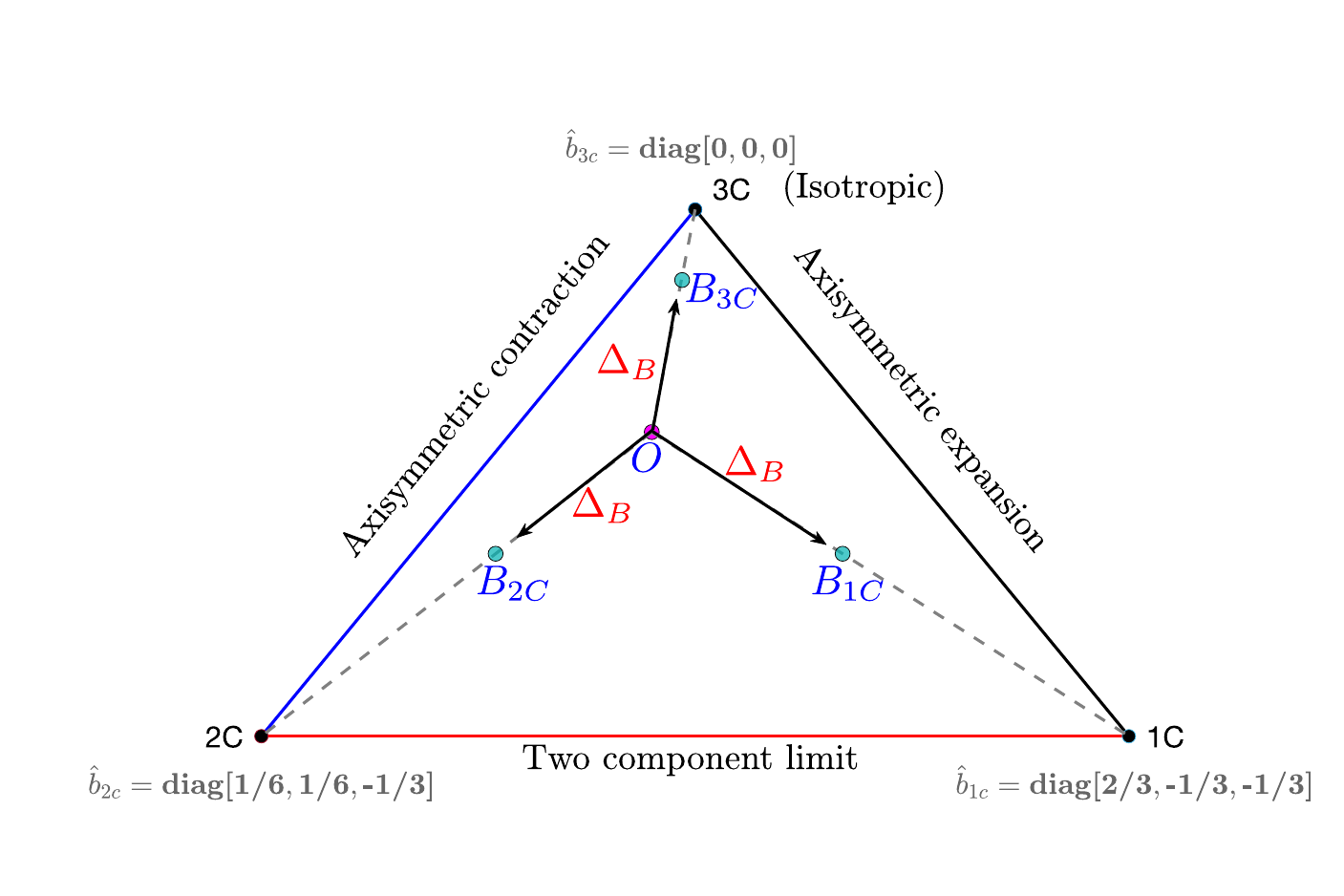}}
\caption{Visualization of the Barycentric Triangle representation for turbulence flow states. Realizable states of the Reynolds stresses are constrained inside or on the edges of the triangle.}
\label{fig:BMap_Sketch.pdf}
\end{figure}

By approximating the unknown Reynolds stresses, RANS modeling attempts to close the time-averaged momentum equations and provides deterministic predictions for flow behaviors under particular circumstances. For turbulence models, the Reynolds stresses $R_{ij}=\langle u_iu_j \rangle$ are a crucial prediction. Reynolds stress decomposed into the anisotropic and isotropic components with

\begin{equation}
R_{ij}=2k(b_{ij}+\frac{\delta_{ij}}{3}),
\end{equation}
where $k(=\frac{R_{ii}}{2})$ is the turbulence kinetic energy, $b_{ij}(=\frac{R_{ij}}{2k}-\frac{\delta_{ij}}{3})$ is the anisotropy tensor. 

Alternatively, the uncertainty intervals can be introduced in place of the unknown stresses to close the time-averaged equations. As a result, potential behaviors are estimated rather than providing precise estimates. Banerjee \textit{et al.} introduced the concept of a two-dimensional barycentric triangle to facilitate visual representation of the realizable Reynolds stress states. This concept is based on the idea of Reynolds stress tensor satisfying physical realizability constraints \cite{schumann1977realizability, lumley1979computational, durbin1994realizability, banerjee2007presentation}. Within this Barycentric map, Reynolds stress states need to be restricted \cite{banerjee2007presentation}. In spectral space the Reynolds stress anisotropy is represented via
\begin{equation}
b_{in}v_{nl}=v_{in}\Lambda_{nl},
\end{equation}
where $v_{nl}$ is a matrix of orthonormal eigenvectors, $\Lambda_{nl}$ is the diagonal matrix of eigenvalues $\lambda_{k}$ and is traceless. Multiplication by $v_{jl}$ gives $b_{ij}=v_{in}\Lambda_{nl}v_{jl}$. This is substituted into the spectral form of the Reynolds stress anisotropy to give
\begin{equation}
R_{ij}=2k(v_{in}\Lambda_{nl}v_{jl}+\frac{\delta_{ij}}{3}).
\end{equation}

According to the order of $v$ and $\Lambda$, $\lambda_{1}\geq\lambda_{2}\geq\lambda_{3}$. Using this, the turbulence anisotropy eigenvalues $\lambda_l$, eigenvectors $v_{ij}$, and the turbulent kinetic energy $k$ reflect the Reynolds stress ellipsoid's form, direction, and amplitude.

The perturbed Reynolds stress tensor can be expressed as
\begin{equation}
R_{ij}^*=2k^* (\frac{\delta_{ij}}{3}+v^*_{in}\Lambda^*_{nl}v^*_{lj})
\end{equation}
where $^*$ represents quantities after perturbation. $k^*=k+\Delta k$ is the perturbed turbulent kinetic energy, $v^*_{in}$ is the perturbed eigenvector matrix, and, $\Lambda^*_{nl}$ is the diagonal matrix of perturbed eigenvalues, $\lambda_l^*$. 

The realizability requirements on $\left\langle u_{i}u_{j} \right\rangle$ for eigenvalue perturbations are enforced by the barycentric map \cite{banerjee2007presentation}, as Fig. \ref{fig:BMap_Sketch.pdf} illustrates. Pecnik and Iaccarino \cite{emory2011modeling} were the ones who first suggested this method. The three extreme states of componentiality of $\left\langle u_{i}u_{j} \right\rangle$ are $1c$, $2c$, and $3c$ on the three corners of the map. Physically, $1c$ denotes a primary fluctuation that is "rod-like" in one direction, $2c$ denotes a principle fluctuation that is "pancake-like" and has the same intensity in two directions, and $3c$ denotes a principal fluctuation that is "spherical" and has the same intensity in three directions. Any realizable $\left\langle u_{i}u_{j} \right\rangle$ may be found via a convex combination of the three vertices $\mathbf{x}_{i c}$ (limiting states) and $\lambda_{l}$ given an arbitrary point $\mathbf{x}$ inside the barycentric map. $\lambda_l^*=B^{-1}\mathbf{x^*}$ yields the altered eigenvalues. $\mathbf{x^*}=\mathbf{x} +\Delta_B(\mathbf{x^t}-\mathbf{x})$ is the representation of the perturbation in the barycentric triangle with $\mathbf{x}$ being the unperturbed state in the barycentric map, $\mathbf{x^*}$ representing the perturbed position, $\mathbf{x^t}$ representing the state perturbed toward and $\Delta_B$ is the magnitude of the perturbation. In this context, $\lambda_l^*=B^{-1}\mathbf{x^*}$ can be simplified to $\lambda_l^*=(1-\Delta_B)\lambda_l + \Delta_B B^{-1}\mathbf{x^t}$. In this case, $B$ creates a linear map from the eigenvalue perturbations to the perturbation in the barycentric triangle.As the goal states, we have $B^{-1}x_{1C} = (2/3, -1/3,-1/3)^T$, $B^{-1}x_{2C} = (1/6, 1/6,-1/3)^T$, and $B^{-1}x_{3C} = (0,0,0)^T$ with the three vertices $x_{1C}$, $x_{2C}$, and $x_{3C}$.

\section{Perturbation of Reynolds Stress Eigenvalues}
Through its Spectral decomposition, the Reynolds stress tensor predicted by the turbulence model is represented by the Eigenvalue perturbation. The Reynolds stress tensor corresponding to this perturbed Spectral decomposition is then reconfigured in physical space after it has disrupted the eigenvalues of this Spectral decomposition. Overall, an ellipsoid may be used to illustrate the expected Reynolds stress tensor of the model. This ellipsoid differs in \textit{shape} due to the Eigenvalue perturbation. 

The anisotropy of turbulent flows cannot be well represented by eddy viscosity-based turbulence models, which is why the Eigenvalue perturbation is required. One line, known as the Plane Strain Line, may be used to represent the Reynolds stress predictions of eddy viscosity-based turbulence models on the Barycentric triangle. The Reynolds stresses in a turbulent flow can sit anywhere inside the Barycentric triangle in real-world turbulent flows. Because of this constraint, turbulence models are unable to accurately depict turbulent flows. A fully developed turbulent pipe flow in a pipe with a non-circular cross-section serves as an example of a flow. These flows exhibit a secondary flow in which the normal stresses in the plane transverse to the streamwise direction differ due to the streamwise mean velocity. This difference in the normal stresses cannot be captured by eddy viscosity based turbulence models, and the secondary flows are not included in the model projections. 

\begin{figure} 
\centerline{\includegraphics[width=\textwidth,trim={0 5.5cm 0 5cm},clip]{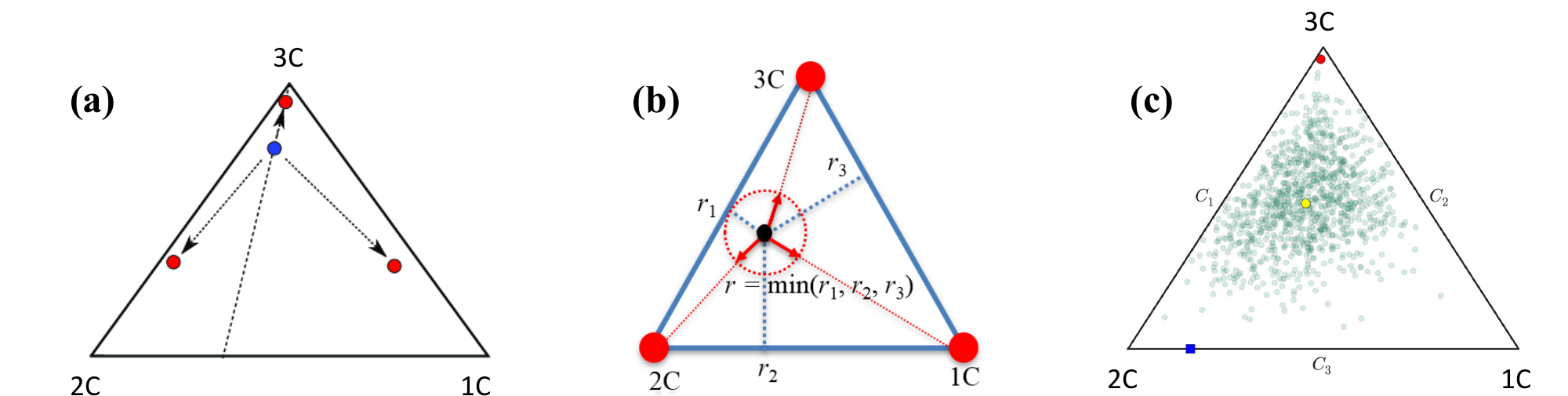}}
\caption{An outline of the various methods that are employed to perturb the eigenvalues. (a) Perturbations are uniformly relative to the limiting states. (b) Perturbations are non-uniform towards the limiting states. (c) The random matrix approach involves eigenvalue perturbations that do not move towards the limiting states.}
\label{fig:eigenvalueperturbations}
\end{figure}

The Barycentric triangle contains all of the turbulence anisotropy realizable states. The eigenvalue perturbations are limited in order to preserve realizability, resulting in the perturbed eigenvalues being mapped to a point inside the Barycentric triangle. The eigenvalue perturbation is solely limited by this. Any framework that satisfies this restriction can theoretically be used to perform these perturbations. There are three approaches that have been used:
\begin{enumerate}
    \item The most often applied method for eigenvalue perturbations is to make changes in the direction of the Barycentric triangle's veterinarianrices. Three perturbed states (in the direction of the 1C, 2C, and 3C states of anisotropy) are adequate when using this method. Either machine learning models or domain knowledge of the turbulent flow can be used to anticipate the size of these disruptions. 
    These disruptions may be non-uniform or uniform. The most prevalent uniform scenario perturbs the 1C, 2C, and 3C states by an equivalent relative size. It is less frequent for there to be an uneven disturbance in the direction of the Barycentric triangle's vertices.  The fundamental concept behind the non-uniform perturbation is that, for every location in the flow domain, the eigenvalue perturbation's amplitude shouldn't be the same in all directions. An alternative is provided by this non-uniform eigenvalue perturbation, in which the amplitude of the perturbation is determined by calculating the distance between the expected Reynolds stresses and the sides of the Barycentric triangle \cite{huangnonuniform}. As a result, each point in the flow domain experiences a disturbance of varying size. 
    
    \item Random perturbations over the Barycentric triangle: Using this method, samples are taken from the probability density function to form an ensemble, which is defined as a probability distribution over the Reynolds stress tensor \cite{xiao2017random}. Using the maximum entropy principle, this is a probabilistic model of a random field of positive semi-definite matrices with a certain mean and correlation structure. Gaussian probability density functions with the covariance function are produced to sample from this random matrix distribution. The space of tensors of the appropriate dimension is mapped to these samples from this probability density function. 
    
    \item Single perturbations toward a rectified state: In the previously stated techniques, the user or a data-driven methodology determines the perturbation's magnitude. Neither the ML algorithm nor the user affect the direction. The disturbances may point randomly or in the direction of the turbulence's limiting states. Several researchers have attempted to forecast the perturbation's path and amplitude using machine learning techniques. The direction and amount of the perturbation indicate the alignment of the disagreement between the high fidelity data (LES/DNS) on the Barycentric triangle and the absolute distance between the predictions of the RANS model. This creates a perturbation vector given the direction and magnitude. The RANS model's predictions can be adjusted with this perturbation vector to get them closer to the eigenvalues found in the high fidelity data. 
\end{enumerate}

A number of researchers have enhanced the eigenvalue perturbations using data-driven machine learning models \cite{chu2023multi, heyse2021data, heyse2021estimating, matha2023evaluation}. However, models used in machine learning often overfit. The field must integrate physics-based data into the machine learning model in order to enhance the generalization of data-driven methodologies. 

Realizability requirements have been used in models based on classical physics to accomplish this \cite{schumann1977realizability, lumley1979computational, du1977realizability}. The eigenvalue perturbations are bounded inside the Barycentric triangle due to realizability restrictions. Even if this restriction is required, it might not be enough. Only the state of the final Reynolds stress tensor is constrained by the confining of the perturbed eigenvalues inside the Barycentric triangle. The modifications produce fresh iterations of the fundamental RANS models, and we must limit the Reynolds stress tensor dynamics that these altered iterations establish. Previous studies have uncovered several restrictions on the Reynolds stress tensor dynamics caused by eigenvalue perturbations. For instance, \cite{mishra2019theoretical} has demonstrated that it is not enough to just confine the perturbed eigenvalues inside the Barycentric triangle.  

A further unanswered question relates to realizability. Although it does not mandate it, the constraint on the eigenvalue perturbations to remain inside the Barycentric triangle preserves the realizability of the projected Reynolds stresses. Unrealizable Reynolds stresses are frequently predicted by turbulence models based on eddy viscosity. Even after the eigenvalue perturbations, these states are unrealisable and are located outside of the Barycentric triangle. While a detailed analysis of the dynamics of realizable Reynolds stresses has been done, it has not been done for these forecasts where the RANS models' initial projections are not achievable. On this subject, more research is definitely required. 

\section{Perturbations to the Reynolds Stress Eigenvectors}
Through its Spectral decomposition, the Eigenvector perturbation reflects the Reynolds stress tensor that the turbulence model predicts. The Reynolds stress tensor corresponding to this perturbed Spectral decomposition is then reconfigured in physical space after it perturbs the eigenvectors of this Spectral decomposition. An ellipsoid can be used to depict the Reynolds stress tensor in general terms. This ellipsoid is altered in terms of \textit{alignment} by the Eigenvector perturbation. 

Since eddy viscosity-based models are unable to predict a Reynolds stress tensor with eigenvectors that deviate from the mean velocity field, the Eigenvector perturbation becomes necessary. The mean rate of strain tensor and Reynolds stresses are related in eddy viscosity models. The eddy viscosity model's projected Reynolds stress tensor is forced to have the same eigenvectors as the mean rate of strain tensor due to the two tensors' linear connection. In turbulent flows found in real life, such as those across complicated surfaces, with streamline curvature, re-attachment, separation bubbles, etc., this is not true. 

There are three broad approaches that have been used for using eigenvector perturbations:
\begin{enumerate}
    \item Eigenvector disturbances to extreme production states: This method focuses on what misalignments are permitted by physics rather than what misalignments between the projected eigen-directions of the RANS model and the actual eigen-directions are likely to occur. This highest misalignment condition is found by using the turbulent production mechanism. The mean rate of strain tensor and the expected Reynolds stress tensor have two alignments that maximize and minimize the formation of turbulence. The projected eigen-directions of the RANS model are altered till these extremal states.

    \item Eigenvector perturbations using Euler bases and incremental rotations: This method concentrates on plausible misaligned states rather than physically feasible ones. The rotational pattern that would match the predicted eigen-directions of the RANS model with the real eigen-directions is often trained into a machine learning model. The Euler angles serve as the reference system for these chained rotations. Starting with the orientation of the anticipated eigen-directions of the RANS model, the orientation of the high fidelity data's eigen-directions may be obtained by applying a certain series of intrinsic rotations, the magnitudes of which are the Euler angles of the desired orientation. The machine learning model is able to learn these Euler angles.

    \item Physical differential equations guiding eigenvector perturbations: In this method, extra differential equations based on physics are utilized to direct the eigenvector perturbations. Finding the eigenvector perturbations that are compatible with the eigenvalue perturbations using the Reynolds Stress Transport equations is a crucial method \cite{thompson2019eigenvector}. The momentum equations of the Reynolds Stress Transport equations employ this Reynolds stress tensor and the eigenvalue perturbations. The right eigenvector perturbations are inferred using the Reynolds stress that is obtained from solving the Reynolds Stress Transport equation. As a result, there is a lack of physical consistency as the eigenvector perturbations rely on the eigenvalue perturbations. 
\end{enumerate}

To regulate the eigenvector perturbations, researchers have employed machine learning models \cite{xiao2017random, wu2018physics}. We must integrate physics-based data into the machine learning model in order to enhance the generalization of data-driven techniques. Unrealisable Reynolds stress tensor values and un-physical Reynolds stress tensor dynamics can result from the eigenvector perturbation without any limitations, as demonstrated by a recent research \cite{matha2023improved, matha2023physically}. When the ultimate dynamics of the eigenspace perturbation diverge from what is predicted based on an intuitive understanding, this also results in a lack of self consistency. They generate and formulate a set of physics-based restrictions that are required in order to achieve a feasible eigenvector perturbation.

\section{Perturbations to the Reynolds Stress Turbulence kinetic energy}
Through its Spectral decomposition, the Reynolds stress tensor anticipated by the turbulence model is represented by the Turbulent Kinetic Energy perturbation. The Reynolds stress tensor corresponding to this perturbed Spectral decomposition is then reconfigured in physical space after it has disrupted the amplitude of this Spectral decomposition. An ellipsoid can be used to depict the Reynolds stress tensor in general terms. The disturbance of turbulent kinetic energy alters this ellipsoid's \textit{size}. 

The incapacity of eddy viscosity based models to represent the physics of turbulence using a scalar isotropic eddy viscosity coefficient gives rise to the necessity of the turbulent kinetic energy perturbation. The ideal value of this coefficient has been observed to vary widely between flows and between domains within the same turbulent flow example in the literature. One may consider the isotropic eddy viscosity coefficient's single scalar value to be a "best-fit" compromise. The eddy viscosity-based model predictions involve significant epistemic uncertainty as a result of this tradeoff. The turbulent kinetic perturbations enable variation in the value of the turbulent eddy viscosity coefficient: $\frac{k}{k^*} = \frac{C^*_{\mu}}{C_{\mu}}$, where $k$ is the turbulent kinetic energy, $C_{\mu}$ is the value of the coefficient of eddy viscosity and starred quantities are the perturbed variants.

As of yet, there are no methods for perturbing the turbulent kinetic energy that are solely based on physics. To alter the turbulent kinetic energy, several researchers have employed Machine Learning models using approximative parameterizations. For instance, the perturbation of $k$ is applied in \cite{cremades2019reynolds} using the parameter $\eta \geq 1$, which specifies the boundaries of the perturbation of turbulent kinetic energy. $k^* = \eta k$ corresponds to the greatest perturbed turbulent kinetic energy, whereas $k^* = k/\eta$ corresponds to the lowest. The machine learning model is used to learn the parameter $\eta$. 

One major drawback of the Eigenspace Perturbation Framework is the lack of a well-established mechanism for the perturbation of the turbulent kinetic energy. Errors in the turbulent kinetic energy can raise prediction disagreement by orders of magnitude more than those in the expected eigenvalues or eigenvectors, when considering the goal of reducing the discrepancy between the RANS model predictions and high fidelity data. Limiting the disrupted turbulent kinetic energy is a hurdle in establishing this turbulent kinetic energy approach. It is obvious that the perturbed turbulent kinetic energy should have a non-negative lower bound. However, for a generic turbulent flow, no upper bound on the turbulent kinetic energy is found. Then the next outstanding question is the question of how to apply these disturbances. $k^* = \eta k$ represents a multiplicative disturbance, whereas $k^* = \delta k + k$ represents an additive perturbation. An investigation of the relative stability of these two strategies would be a significant step forward in the use of data-driven strategies for perturbation methods used to quantify uncertainty in RANS models.

Although the three main forms of perturbations—that is, to the turbulent kinetic energy, eigenvalues, and eigenvectors—have been covered individually, they are applied in tandem with one another. Therefore, even if realizability criteria for every kind of perturbation could be required, they might not be adequate in the overall scenario where every kind of perturbation is used. Therefore, it is crucial to create realizability restrictions for the typical scenario in which each of the three perturbation types is used. Furthermore, despite the fact that research has concentrated on maintaining the realizability of the RANS forecasts of the Reynolds stresses, it has overlooked the reality that RANS predictions are frequently not achievable. Both an unrealizable and a realizable Reynolds stress tensor are maintained by the Eigenspace Perturbation Method in order for it to stay unrealizable. Even if the only purpose of the study is to determine if the perturbations are making these points more unrealizable, the impacts of the Eigenspace Perturbation Method based perturbations on the unrealizable predictions from the RANS models need to be examined.

\section{Summary, Conclusions and Future Directions}
In engineering design and analysis challenges, turbulent fluxes are significant. Computational fluid dynamics research of turbulent flows currently and in the future will be dependent on turbulence models in large eddy simulations and Reynolds Averaged Navier Stokes modeling. These simulations based on turbulence models suffer from several sources of uncertainty. We require trustworthy estimations of these predicted uncertainty in safety-critical engineering design applications. For CFD simulations, this type of uncertainty quantification is our main emphasis \cite{karniadakis1995toward, cavallo2007error}. We examine the latest developments in the estimation of various uncertainty components, such as the aleatoric\cite{smith2013uncertainty}, numerical \cite{denton2010some, coleman1997uncertainties, katz2011mesh}, and epistemic components\cite{iaccarino2017eigenspace}. We describe in further detail how these uncertainties might be estimated using Machine Learning (ML) techniques. Above all, we describe specific shortcomings in these methods. We describe the necessity of a framework for the perturbation of the turbulent kinetic energy grounded on physics. In order to guarantee the physical consistency of the perturbation, eigenvalues, eigenvectors, and turbulent kinetic energy, we additionally emphasize the necessity of realizability criteria. They comprise the requirements for both adequate and essential realizability. We draw attention to the fact that although modifications to the simulated Reynolds stress preserve a feasible state for the Reynolds stress, their impact on the turbulence model's unrealistic predictions remains unstudied. Although a lot of research has been done on the application of machine learning models for uncertainty estimation, these models still need to be interpretable. It is also crucial to include physics information into Machine Learning models for CFD UQ. Drawing upon this research, we identify key issues that require attention and provide targeted measures to overcome these constraints.



 \bibliographystyle{elsarticle-num} 
 \bibliography{cas-refs}





\end{document}